\documentclass[12pt]{article}
\usepackage{graphicx, amsmath, amssymb, cite, setspace, color}

\usepackage[top=1 in, bottom=1 in, left=1 in, right=1 in]{geometry}

\newcommand{\captionfonts}{\footnotesize} 
\makeatletter 
\long\def\@makecaption#1#2{%
  \vskip\abovecaptionskip
  \sbox\@tempboxa{{\captionfonts #1: #2}}%
  \ifdim \wd\@tempboxa >\hsize
    {\captionfonts #1: #2\par}
  \else
    \hbox to\hsize{\hfil\box\@tempboxa\hfil}%
  \fi
  \vskip\belowcaptionskip}
\makeatother 

\def\fnote#1#2{\begingroup\def\thefootnote{#1}\footnote{#2}
     \addtocounter{footnote}{-1}\endgroup}

\begin{document}
\title{On `Nothing'}

\author{Adam~R.~Brown$^{1,\, 2}$ and Alex~Dahlen$^{1,\,3}$ \vspace{.1 in}\\
 \vspace{-.3 em}  $^1$ \textit{\small{Physics Department, Princeton University, Princeton, NJ 08544, USA}} \\
 \vspace{-.3 em}  $^2$ \textit{\small{Princeton Center for Theoretical Science, Princeton, NJ 08544, USA}} \\
 \vspace{-.3 em}  $^3$ \textit{\small{Berkeley Center for Theoretical Physics, Berkeley, CA 94720, USA}} }
\date{}

\maketitle
\fnote{}{\hspace{-.65cm}emails: \tt{adambro@princeton.edu, adahlen@berkeley.edu}}
\vspace{-.95cm}

\begin{abstract}

\noindent Nothing---the absence of spacetime---can be either an endpoint of tunneling, as in the bubble of nothing, or a starting point for tunneling, as in the quantum creation of a universe. We argue that these two tunnelings can be treated within a unified framework, and that, in both cases, nothing should be thought of as the limit of anti-de Sitter space in which the curvature length approaches zero.
To study nothing, we study decays in models with perturbatively stabilized extra dimensions, which admit not just bubbles of nothing---topology-changing transitions in which the extra dimensions pinch off and a hole forms in spacetime---but also a whole family of topology-preserving transitions that nonetheless smoothly hollow out and approach the bubble of nothing in one limit. The bubble solutions that are close to this limit, bubbles of next-to-nothing, give us a controlled setting in which to understand nothing.
Armed with this understanding, we are able to embed proposed mechanisms for the reverse process, tunneling from nothing to something, within the relatively secure foundation of the Coleman-De Luccia formalism and show that the Hawking-Turok instanton does not mediate the quantum creation of a universe.
\vspace*{6 cm}
\end{abstract}

\pagebreak

\section{Introduction}

`Nothing' first made an appearance in modern physics with the work of Witten \cite{Witten:1981gi}, who showed that spacetimes with compact extra dimensions can be unstable to decay---the compact extra dimensions can pinch off to form a Ôbubble of nothingÕ containing not only no matter and no fields but also no space and no time. ÊIf the universe can tunnel to nothing, it is natural to ask whether it can tunnel from nothing---the quantum creation of a universe.Ê Several authors have addressed this question \cite{Vilenkin:1983xq,Linde:1983mx,Hartle:1983ai,Hawking:1998bn,BlancoPillado:2011me}, but unfortunately the situation remains somewhat murky. At least part of this murkiness stems from the ambiguity of what is meant by `nothing'.   In this paper, we address this ambiguity.

Our first step is to ask what the decay of Kaluza-Klein spacetime tells us about nothing. Unfortunately the bubble of nothing itself turns out to be somewhat enigmatic on this question.  However, in Sec.~\ref{sec:sec2}, we show that in models with perturbatively stabilized extra dimensions, bubbles of nothing are not all we have to work with.   The bubble of nothing is not an isolated solution, instead these models admit whole families of possible decays which remove different amounts of the stabilizing potential.   Only the decay that removes all the stabilizing potential, the bubble of nothing, changes the topology of spacetime; the other tunneling solutions in the family are topology-preserving but they nonetheless smoothly approach the bubble of nothing in the limit in which all the stabilizing potential is removed.  Though we learn little from  the bubble of nothing itself, we learn a great deal from the sequence of bubbles that approach it, the bubbles of next-to-nothing. 

Two things happen to the bubble interior as the bubble of nothing is approached: the extra dimensions shrink to zero size and smoothly pinch off; and  simultaneously the effective potential becomes ever more negative. From the perspective of the lower-dimension Einstein frame, therefore,  the interior becomes more and more negatively curved, so that
\begin{quote} 
\emph{`Nothing' should be thought of as the limit of anti-de Sitter space in which the curvature length goes to zero.}
\end{quote}
As it goes deeper and deeper into anti-de Sitter space, the interior of the bubble empties out until nothing remains. Ê

In Sec.~\ref{sec:sec3} we turn to the quantum creation of the universe. Quantum transitions in dynamical spacetimes are described by the Coleman-De Luccia formalism \cite{Coleman:1980aw}. This formalism is  well studied, well understood, and is the most credible part of semiclassical quantum gravity.  While it is always used in the context of transitions to nothing, it is not always used for transitions from nothing, perhaps because of the ambiguity of how to treat nothing.  With our new understanding of nothing, however,  we are able to embed otherwise poorly moored questions of the quantum creation of the universe within this secure foundation.

The Coleman-De Luccia formalism requires that the same instanton that describes tunneling to nothing should also describe tunneling from nothing.  Within the context of this unified framework, we should think of tunneling from something to nothing as down-tunneling, and tunneling from nothing to something as up-tunneling.  The `nothing's in the two processes are therefore the same---they are both the limit of anti-de Sitter space as the curvature length goes to zero.  But, since up-tunneling from anti-de Sitter space is impossible, the quantum creation of a universe using a bubble of nothing instanton---the Hawking-Turok process \cite{Hawking:1998bn}---is forbidden.

\section{`Nothing' as the limit of anti-de Sitter} \label{sec:sec2}

In this section we study `nothing' using the bubble of nothing.  In Sec.~\ref{sec:wittenbon} we review the bubble of nothing in its original context, that of a single unstabilized extra dimension.  In Sec.~\ref{sec:fluxbon} we study the approach to the bubble of nothing in a simple model, the 6D Einstein-Maxwell theory, and use it to draw conclusions about nothing.  In Sec.~\ref{sec:bongeneral} we argue that these conclusions are not special to the 6D Einstein-Maxwell model but are in fact  general.

\subsection{Bubbles of nothing with unstabilized extra dimensions} \label{sec:wittenbon}

\begin{figure}[htbp] 
   \centering
   \includegraphics[width=6in]{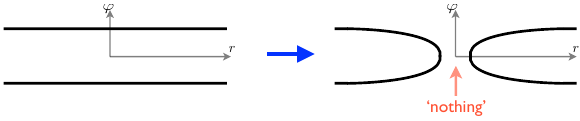}  
   \caption{On the left: a cross-section through Minkowski$_{4} \times S^1$. The uncompactified dimensions are aligned horizontally, the extra dimension is vertical. The top line (the plane $\varphi=\pi$) and the bottom line (the plane $\varphi = - \pi$) are identified, so the extra dimension is periodic. On the right: cross-section through the center of the  bubble of nothing at the instant of nucleation. At large area-radius $r \gg R$, the space is barely perturbed from the original Minkowski$_4 \times S^1$. As you approach the center, the size of the extra dimension shrinks until, at $r=R$, spacetime smoothly pinches off.} 
   \label{fig:BubOfNothprofile}
\end{figure}

Consider 3+1+1-dimensional Minkowski spacetime with the extra dimension compactified on a circle, so that the metric is
\begin{equation}
ds^2=-dt^2+dx^2+ dy^2 + dz^2+R^2 d\varphi^2, 
\label{McrossS}
\end{equation}
where $\varphi$ is periodically identified with period $2\pi$, so that the size of the extra dimension is $2 \pi R$.  This size is unstabilized---the four-dimensional effective theory contains a massless radion field---but this spacetime is otherwise classically stable.  Quantum mechanically, however, this spacetime is unstable to the nucleation of a bubble of nothing \cite{Witten:1981gi}.  Decay is mediated by a Euclidean instanton given by
\begin{equation}
ds^2 = \frac{dr^2}{1 - R^2 /r^{2}} + r^2 \left( d\theta^{\,2} + \cos ^2 \theta \, d \Omega_2^{\, 2} \right)  + \left( 1 - {R^2 /r^{2}} \right) R^2 d\varphi^2, \label{eq:Wittencoordinates}
\end{equation}
 with $R \le r<\infty$ and $\varphi$ still periodically identified. Figure~\ref{fig:BubOfNothprofile} shows the transition mediated by this instanton.  In the semiclassical description, the spatial metric makes a quantum jump from  a fixed $t$-slice of the metric in Eq.~\ref{McrossS} to the $\theta=0$ slice of  the metric in Eq.~\ref{eq:Wittencoordinates}. 
Far from the center of the bubble ($r \gg R$), the size of the circle remains $2 \pi R$. Closer to the center of the bubble, the circle shrinks until at $r=R$, the size of the extra dimension goes smoothly to zero and spacetime pinches off. A three-dimensional slice through the bubble now has a hole. A sphere that surrounds this hole cannot have surface area less than $4 \pi R^2$, it can contract to $r=R$ but no further. Despite the pinch-off, there is no singularity and indeed the curvature is nowhere large, so the semiclassical approximation remains reliable.

\begin{figure}[htbp] 
   \centering
   \includegraphics[width=3.9in]{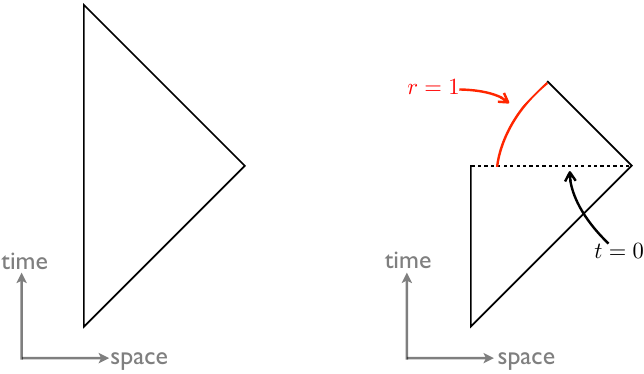} 
   \caption{On the left: the Penrose diagram for the original 3+1+1-dimensional Minkowski spacetime, with the extra dimension, as well as the angular dimensions, suppressed. On the right: the semiclassical description of the formation and growth of a bubble of nothing. Along the dotted line at $t=0$, the spacetime makes a quantum jump: before, the spacetime has the metric of Eq.~\ref{McrossS}; after, the spacetime has the metric of Eq.~\ref{eq:movingbon}.  The extra dimension degenerates to zero size at $r=R$; for $r<R$ there is nothing. The bubble wall then expands outward and the hole in spacetime grows.}
   \label{fig:example}
\end{figure}

After nucleation, the bubble begins expanding.  Its subsequent evolution, shown in Fig.~\ref{fig:example}, can be found by analytically continuing  $\theta \rightarrow i t$ in Eq.~\ref{eq:Wittencoordinates} to give
\begin{equation}
ds^2 = \frac{dr^2}{1 - R^2/r^{2}} + r^2 (-dt^2 + \cosh^2 t \, d \Omega_2^{\,2} ) + \left( 1 - {R^2/r^{2}} \right) R^2 d\varphi^2. \label{eq:movingbon}
\end{equation}
These coordinates slice spacetime with time-like hyperbolae, so in ordinary circumstances they would miss the inside of the lightcone. In this case, however, there is no inside of the lightcone, because spacetime only exists for $r\ge R$.  The bubble of nothing starts at rest, but accelerates out, so that the minimal-area sphere containing the bubble has area  $4 \pi R^2 \cosh^2 t$.  Particles that approach the edge of the bubble go around the lip and are boosted out again, albeit on the other side of the extra dimension $\varphi \rightarrow \varphi + \pi$. (Eventually, because it is accelerating, the wall will catch up with the particle and the collision sequence will repeat.  In the instantaneous local rest frame the infinite sequence of collisions look exactly alike: they are separated by the same proper time and the particle reaches the same maximum proper distance from the bubble wall.) 

The bubble of nothing is an insatiable wall of annihilation, advancing unremittingly, ever closer to the speed of light, and leaving nothing in its wake.

\subsection{Bubbles of nothing in stabilized extra dimensions: \\ 
a simple example} \label{sec:fluxbon}

In the last subsection, we described the bubble of nothing in the context of an extra dimension that was unstabilized.  This bubble of nothing instanton is an isolated solution, in the sense that it is the only possible decay. In this subsection, we will describe the bubble of nothing instanton in the context of extra dimensions that are perturbatively stabilized. In this case, the bubble of nothing is no longer isolated: as well as the bubble of nothing, there is now a whole family of possible decays that do not change the spacetime topology, but that nonetheless smoothly approach the bubble of nothing---a family of bubbles with a limit in which spacetime hollows out. By examining solutions close to the bubble of nothing, we will gain insight into the nature of nothing.

Let's begin by studying this approach to the bubble of nothing in the simplest possible model with perturbatively stabilized extra dimensions, the 6D Einstein-Maxwell theory \cite{Freund:1980xh,BlancoPillado:2009di}, a model whose tunneling instantons we constructed explicitly in \cite{Brown:2010mf}.

\subsubsection{Introduction to the 6D Einstein-Maxwell theory}

The action for the 6D Einstein-Maxwell theory is 
\begin{equation}
S_\text{EM}= \int d^{\,6}\!\,x \, \sqrt{-G}\left(\frac{M_6^{\,4}}{2} \mathcal{R}^{(6)}-\frac14 F_{AB}F^{AB}-\Lambda_6\right), \label{eq:EMaction}
\end{equation}
where $F_{AB}$ is the electro-magnetic field strength,  $\Lambda_6$ is a positive six-dimensional cosmological constant, and $M_6$ is the 6D Planck mass.  
We will examine the sector of the theory in which two of the dimensions are compactified on a sphere, leaving $3+1$ dimensions large. 

This theory can be re-expressed as an effective four-dimensional theory, at the cost of introducing additional Kaluza-Klein fields. This is done in two steps. The first step is to perform the integral over the two extra dimensions in Eq.~\ref{eq:EMaction}, leaving a four-dimensional action. In this action the effective four-dimensional Planck mass $M_4$ is given by 
\begin{equation}
M_4^{\, 2} = 4 \pi R^2 M_6^{\, 4},
\end{equation}
where $R$ is the size of the extra dimensions.  Measured in units of $M_6$, the four-dimensional Planck mass $M_4$ varies depending on the size of the extra dimensions $R$. From the four-dimensional perspective, however, it is more natural to measure the potential in units of $M_4$. The second step, therefore, is to change coordinates, which is to say to conformally rescale, so that in our new units $M_4$ is independent of $R$: this is called 4D  \emph{Einstein frame}. Einstein frame is the natural choice  from the point of view of four-dimensional observers, and from the point of view of the Coleman-De Luccia tunneling results that we study in Sec.~3. 

\begin{figure}[htbp] 
   \centering
   \includegraphics[width=6.4in]{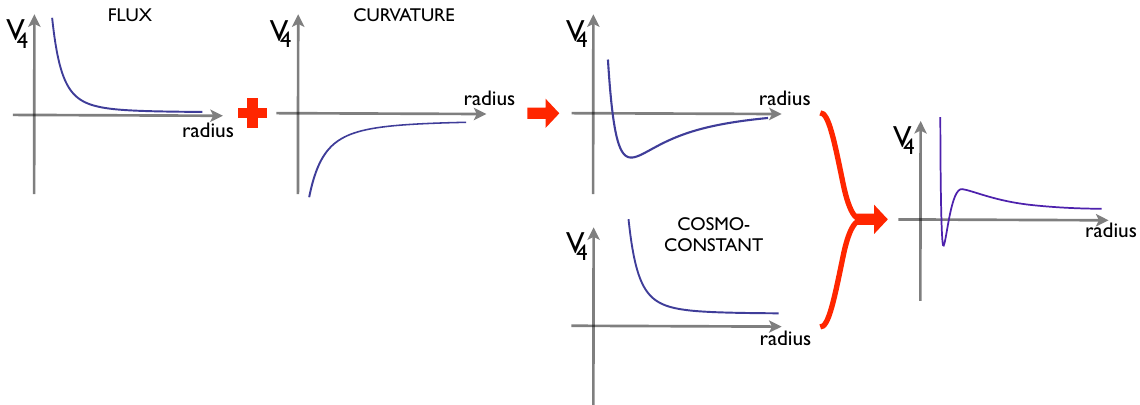} 
   \caption{The combination of the repulsive flux term and the attractive curvature term gives rise to a stable minimum of the radion potential; this minimum necessarily has $V_4<0$.  The fundamental 6D cosmological constant raises this minimum; vacua with enough flux can even get raised up to $V_4>0$.}
   \label{stableminimum}
\end{figure}

In 4D Einstein frame, the size of the extra dimensions becomes a dynamical field. If the two legs of the  flux are wrapped around the two extra dimensions, the effective potential is 
\begin{equation}
\frac{V_4(R)}{M_4^{\, 4}} = \frac{1}{4 \pi} \left( \frac{1}{32 \pi^2} \frac{g^2 N^2}{M_6^{\,2}} \frac{M_6^{\,-6}}{R^6} - \frac{M_6^{\,-4}}{R^4} + \frac{\Lambda_6}{M_6^{\, 6} } \frac{M_6^{\,-2}}{R^2} \right) , \label{eq:V4radion}
\end{equation}
where $N$ is the number of units of flux, and $g$ is the fundamental magnetic charge. 
The sphere's positive curvature makes the sphere want to contract to zero size, but the flux lines resist being squeezed together and therefore buttress the extra dimensions against collapse. As shown in Fig.~\ref{stableminimum}, the combination of these two terms gives rise to a minimum of the radion potential, and so stabilizes the size of the extra dimensions. 
The effect of the cosmological constant is to raise this minimum.

\begin{figure}[htbp] 
   \centering
   \includegraphics[width=4in]{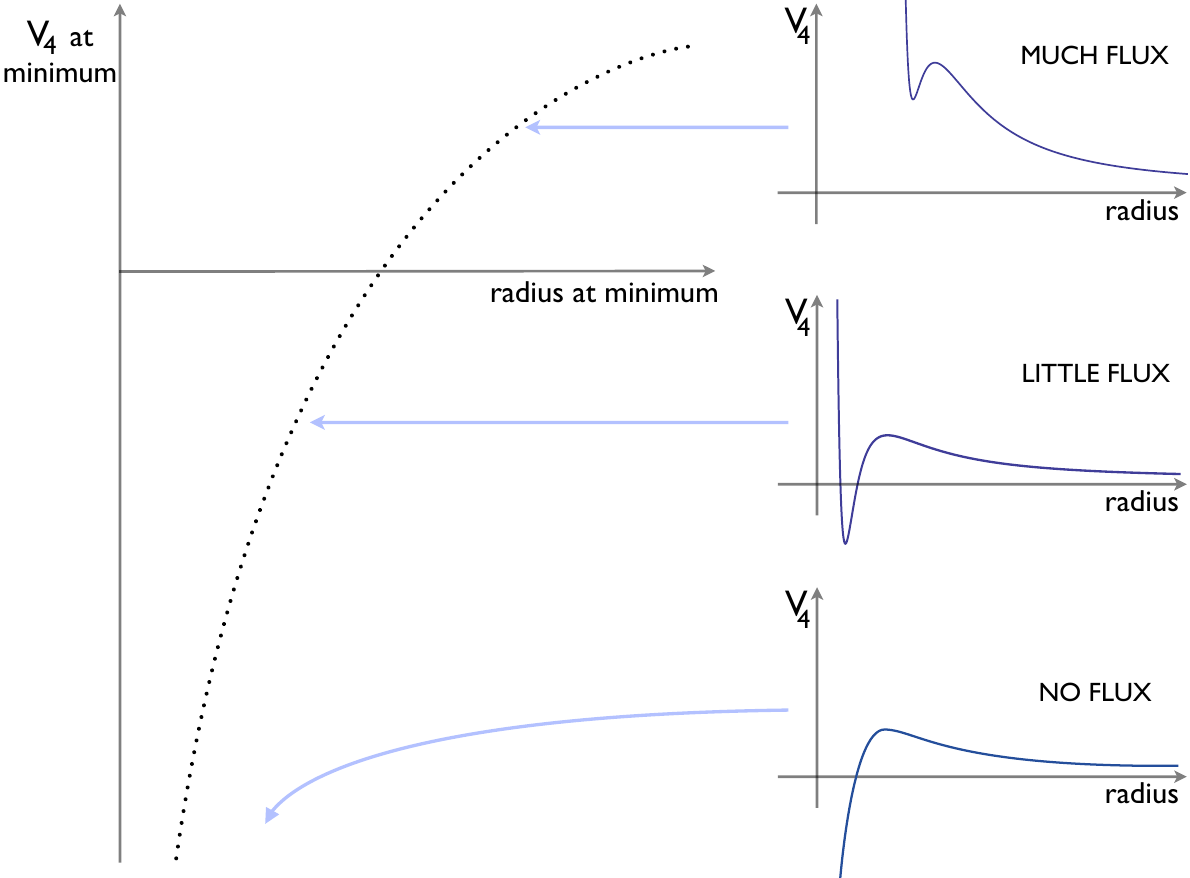}  
   \caption{For a given $\Lambda_6$, varying the number of units of flux wrapping the extra dimensions changes the radion potential.  On the right: three radion potentials are shown with different amounts of flux.  When there are many units of flux, the minimum is at $V_4>0$ and large $R$.  When there are fewer units of flux, the minimum is at $V_4<0$ and small $R$.  When there is no flux, the minimum disappears to $V_4 \rightarrow - \infty$ and $R\rightarrow0$.  (De Sitter vacua have an additional nonperturbative instability, which is to decompactification---the radion tunnels out to larger values. This instability is not relevant to our discussion about nothing.) On the left: the values of $V_4$ and $R$ in the compactified vacua are shown for various values of $N$. The vacua can be made arbitrarily dense by taking the fundamental magnetic quantum $g$ arbitrarily small.}
   \label{fig-radionpotentials}
\end{figure}

Because the flux is quantized, this gives rise to a discrete landscape of vacua. Each vacuum in this landscape has a different integer number of flux units wrapping the extra dimensions, and correspondingly a different radion potential, as shown in Fig.~\ref{fig-radionpotentials}.  Vacua with many units of flux are stabilized at large $R$ and positive $V_4$. The fewer units of flux, the smaller the stable value of $R$ and the less positive the stable value of $V_4$.  The conformal rescaling ensures that in Einstein frame, these vacua all have the same value of $M_4$.  By taking the magnetic charge $g$ to be small, the set of vacua becomes increasingly dense. These vacua are perturbatively stable, but can be nonperturbatively unstable to discharging flux.   
\pagebreak

\vspace{-.1in}

\subsubsection{Flux tunneling in the 6D Einstein-Maxwell theory}
\begin{figure}[htbp] 
   \centering
   \includegraphics[width=5.5in]{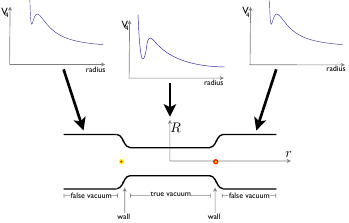} 
\caption{A cross-section through a flux-tunneling instanton.  As in Fig.~\ref{fig:BubOfNothprofile}, extended directions are aligned horizontally and the extra dimensions are aligned vertically. The three-sphere of charged brane, indicated by the colored dots, forms a bubble in the extended dimensions and sits at a point in the extra dimensions. Outside the bubble, there are many units of flux wrapping the extra dimensions. Inside the bubble there are fewer units of flux, because some have been discharged by the brane. This means that the radion inside the bubble lives in a different radion potential, one with a minimum at a smaller value of $R$ and a less positive value of $V_4$. }
   \label{fig-nexttonothing}
\end{figure}

The flux wrapping the extra dimensions is discharged by the quantum nucleation of a charged brane that forms a sphere in the extended directions---a bubble---and sits at a point in the extra dimensions, as in Fig.~\ref{fig-nexttonothing}.  Outside this bubble, the flux is unchanged, but inside the bubble the flux is reduced. The radion potential inside is correspondingly adjusted, and the radion settles into its new minimum in which both $R$ and $V_4$ are less positive.   After nucleation, the bubble classically expands, accelerating toward the speed of light.

\begin{figure}[htbp] 
   \centering
   \includegraphics[width=5.5in]{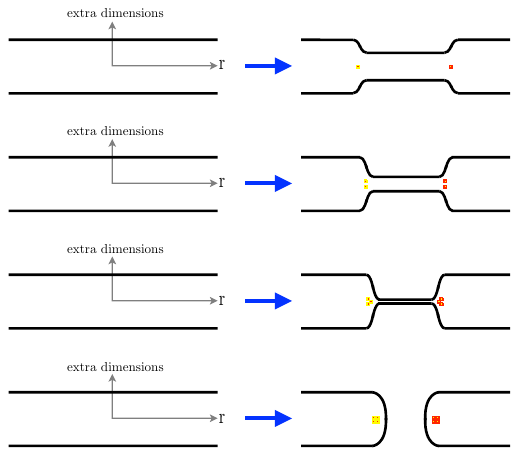}
   \caption{A sequence of tunneling instantons that discharge different amounts of flux. The more charged branes in the stack, the more units of flux are discharged, and the smaller the size of the extra dimensions inside the bubble. In the limit that all the flux is discharged, the area-radius of the bubble stays nonzero, but the size of the extra dimensions inside the bubble goes to zero. This is the bubble of nothing, and can be compared with Fig.~\ref{fig:BubOfNothprofile}. The exact instanton profiles are computed numerically in \cite{Brown:2010mf}; this figure shows the qualitative behavior.}
   \label{fig:fluxsequence}
\end{figure}

For a given vacuum, there is a whole family of possible flux decays; the vacuum can nucleate any number of charged branes, and therefore discharge any number of units of flux inside the bubble.   As shown in Fig.~\ref{fig:fluxsequence}, the larger the stack, the less flux left inside the bubble and the smaller the size of the extra dimensions. The largest stack discharges \emph{all} of the flux, leaving no force to buttress the extra dimensions against collapse; the extra dimensions shrink to zero size and smoothly pinch off, as in a bubble of nothing.\footnote{Note that the charged brane sits in the middle of the wall region, set back from the true vacuum, as depicted in Fig.~\ref{fig-nexttonothing}; its position can be calculated in detail, as in \cite{Brown:2010mf}. Our argument in this paper will only rely on the possibility of nucleating a bubble of nothing, but in fact it can be shown that for some landscapes this is  not only a possible decay, it is the fastest decay---the larger the stack of charged branes, the easier it is to nucleate \cite{Brown:2010mf, Brown:2010bc}.}

\pagebreak

We have thus constructed a whole family of possible decays that do not change the spacetime topology, but that hollow out in one limit. To study nothing, we can now study  members of this family that are close to the bubble of nothing, but that are nevertheless topology-preserving.  In the limit in which there are very few units of flux left in the interior ($g^2N^2/M_6^{\,2}\ll M_6^{\,6}/\Lambda_6$), the cosmological constant in Eq.~\ref{eq:V4radion} becomes subleading and the radius and potential inside the bubble are well approximated by
\begin{eqnarray}
M_6 R&=& \frac{\sqrt{3}}{8 \pi} \,\,\frac{gN}{M_6} \,\,\left[ 1  + O\left(\frac{g^2 N^2}{M_6^{\,2}}\frac{\Lambda_6}{M_6^{\,6}}\right) \right], \\
\frac{V_4}{M_4^{\, 4}}&=&-\frac{1024 \pi^3}{27}\,\,\frac{M_6^{\,4}}{g^4N^4}\,\, \left[ 1  + O\left(\frac{g^2 N^2}{M_6^{\,2}}\frac{\Lambda_6}{M_6^{\,6}}\right) \right].
\end{eqnarray}

As ever more flux is discharged, the size of the extra dimensions inside the bubble becomes ever smaller and the effective potential becomes ever more negative. When $gN=0$ we have the bubble of nothing, and the interior is `nothing'. Therefore, from the four-dimensional perspective, `nothing' corresponds to the limit in which $V_4 \rightarrow - \infty$. In other words, \emph{`nothing' should be thought of as the limit of anti-de Sitter space in which the curvature length goes to zero.} 

This is the central claim of this paper. We spend the remainder of this section justifying this claim, first by showing that this limit is smooth, and, second by showing that this is not a special feature of the 6D Einstein-Maxwell theory and in fact holds in all compactifications that admit a bubble of nothing.

\subsubsection{Bubbles of next-to-nothing} \label{sec:smoothbon}

The bubble of nothing is the limit of flux tunneling in which all the flux is discharged; despite being topology-changing, it is the limit of a family of transitions that are topology-preserving. 
In what sense is this limit smooth? 


For thin branes, the flux tunneling instantons amongst the vacua of the 6D Einstein-Maxwell model break up into three parts: a false-vacuum exterior, a true-vacuum interior, and a brief transitionary wall containing the charged brane, as shown in Fig.~\ref{fig-nexttonothing}.  For decays that leave only a few units of flux inside the bubble, the (Euclidean) interior metric is well approximated by
\begin{eqnarray}
ds^2 & = & \ell^2 \left( d\xi^2 + \sinh^2 \xi d \Omega_3^{\,2} \right) + \frac{1}{6} \ell^2 d \Omega_2^{\,2}, \ \ \ \  \ \textrm{ with } 0 < \xi < \textrm{arcsinh}\, r_0/\ell , \  \textrm{ or} \\
ds^2 & = & \frac{dr^2}{1 + r^2 /\ell^2}  +  r^2 d \Omega_3^{\,2} + \frac{1}{6} \ell^2 d \Omega_2^{\,2}, \ \ \ \ \ \, \ \ \ \textrm{ with } 0<r<r_0,  \label{eq:smallFmetric}
\end{eqnarray}
where $\ell = 3 g N /(\sqrt{32} \pi M_6^{\,2}) =\sqrt6\, R$. As we approach the limit, two-dimensional slices at a fixed point in the extended directions are becoming increasingly positively curved and four-dimensional slices at a fixed point in the compact directions are becoming increasingly negatively curved.\footnote{For the 6D Einstein-Maxwell model, there is a simple way to see how these two divergences are linked. The contribution of flux to the six-dimensional energy momentum tensor is traceless, so the only contribution to the six-dimensional curvature $\mathcal{R}_6$ is from  the bulk cosmological constant; this means that $\mathcal{R}_6$ is fixed, finite and the same for all the decay instantons.  Therefore,  since $\mathcal{R}_6=\mathcal{R}_2+\mathcal{R}_4$, if the size of the extra dimensions goes to zero (sending $\mathcal{R}_2\rightarrow\infty$) then the anti-de Sitter curvature length must also go to zero (sending $\mathcal{R}_4\rightarrow - \infty$). [Notice that $\mathcal{R}_4$ is the curvature of a four-dimensional spatial slice and not the Einstein-frame effective curvature; they differ because of the conformal rescaling. The curvature length in 4D Einstein frame is $\ell_\textrm{curv} \sim g^2 N^2$, whereas the curvature length of a four-dimensional slice is $\ell \sim g N$.] } 
This interior metric is cut off at an area-radius $r_0$ and glued into the wall region; as we showed in \cite{Brown:2010mf}, $r_0$  stays nonzero as $\ell$ goes to zero. 

The question of the smoothness of this limit is the question of to what extent the interior metric of Eq.~\ref{eq:smallFmetric} smoothly approaches nothing. The limit is not \emph{topologically} smooth, but we  now show that it is \emph{geometrically} smooth in three physically important senses:
\begin{enumerate}
\item {\bf Six-dimensional volume smoothly vanishes.} The six-dimensional volume of the true-vacuum region shrinks smoothly to zero as we take $\ell \rightarrow 0$ because the size of the extra dimensions smoothly vanishes,  
\begin{equation}
(\textrm{Volume})_6 = (\textrm{Volume})_2 \times (\textrm{Volume})_4 = \frac{4 \pi \ell^2}{6}  \times (\textrm{Volume})_4 \rightarrow  0. 
\end{equation}
\end{enumerate}

\noindent But it's more than that. Another example of a metric for which the six-dimensional volume inside the bubble smoothly vanishes is
\begin{equation}
ds_{\textrm{not smooth}}^2 = dr^2 + r^2 d \Omega_3^{\,2}  + \ell^2 d \Omega_2^{\,2}, \ \ \ \ \ \ \textrm{ with } 0<r<r_0. \label{eq:notsmooth1}
\end{equation}
For this metric, as $\ell \rightarrow 0$ the extra dimensions shrink to zero size but we are left with a wafer-thin interior region with four-volume but no thickness---the spacetime inside the bubble is flattened but not annihilated. However, the interior metric of flux-tunneling instantons, Eq.~\ref{eq:smallFmetric}, approaches nothing more smoothly than  the metric of Eq.~\ref{eq:notsmooth1}, in the following sense.

\pagebreak

\begin{enumerate}
\item[2.] { \bf Four-dimensional volume smoothly vanishes.} 
Consider a four-dimensional slice through the bubble at a fixed value of the extra dimensions. The four-dimensional volume of this region is 
\begin{equation}
(\textrm{Volume})_4 = 
\int d\Omega_3 \int_0^{r_0} dr \frac{r^3}{\sqrt{1 +r^2 / \ell^2} } = \frac{2 \pi^2}{3} r_0^{\, 3} \ell + O(r_0^{\,2}\ell^2) \rightarrow 0. \label{eq:fourvolume}
\end{equation}
\end{enumerate}
The reason that the four-volume of this slice goes to zero is that the slice is becoming increasingly negatively curved.  Negatively curved spaces 
have the property that, for a given surface area, spheres have less volume than in flat space. Most of the volume lies within a proper distance $\ell$ of the surface; the volume inside of a four-sphere of radius $r$ scales not as $r^4$ but as $r^3\ell$, as is evident from Eq.~\ref{eq:fourvolume}. In the limit that $\ell \rightarrow0$ the volume of any region inside the bubble goes to zero while the surface area stays finite.

\begin{figure}[htbp] 
   \centering
   \includegraphics[width=5.2in]{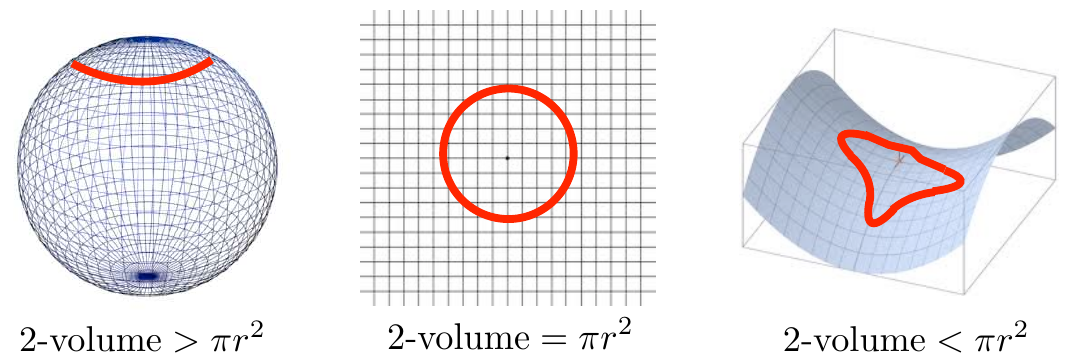} 
   \caption{For a given surface area, spheres in positively curved spaces have more volume than in flat space. Spheres in negatively curved spaces have less volume than in flat space.}
   \label{fig:negativecurvature}
\end{figure}

But it's even more than that. Another example of a metric for which both the six-dimensional volume and the four-dimensional volume inside the bubble smoothly vanish is
\begin{equation}
ds_{\textrm{not smooth}}^2 = \ell^2 dr^2 + r^2 d \Omega_3^{\, 2}  + \ell^2 d \Omega_2^{\, 2}, \ \ \ \ \ \ \textrm{ with } 0<r<r_0.
\label{eq:notsmooth2}
\end{equation}
Consider foliating this metric with spheres labeled by $r$; though these spheres have zero volume as we approach the limit, they have nonzero surface area. When the spheres disappear, they are inside the bubble in the sense that though the proper distance of a sphere from the edge of the bubble goes to zero, the proper distance from the center of the bubble also goes to zero, so the sphere stays a fixed proportion of the way out of the bubble. Therefore, when we reach the limit, the foliating spheres disappear abruptly.  However, the interior metric of flux-tunneling instantons, Eq.~\ref{eq:smallFmetric}, approaches nothing more smoothly than  the metric of Eq.~\ref{eq:notsmooth2}---the foliating spheres do not abruptly disappear, instead they get smoothly ejected. 

\pagebreak

\begin{figure}[htbp] 
   \centering
   \includegraphics[width=3.4in]{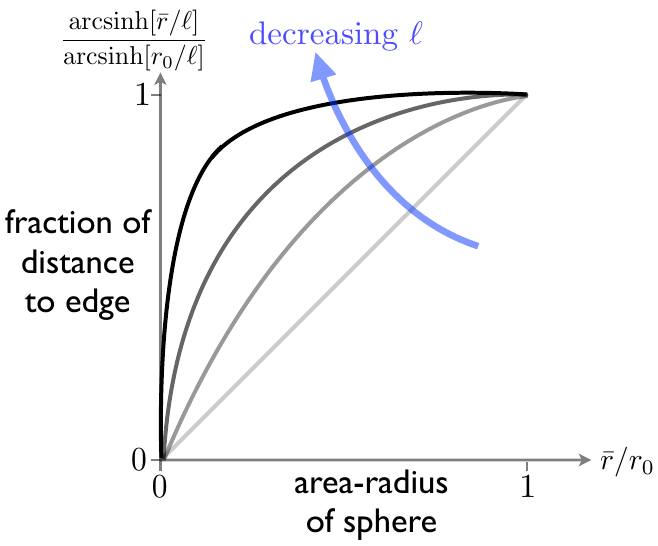} 
   \caption{Consider foliating the interior metric, Eq.~\ref{eq:smallFmetric}, with three-spheres. A sphere of area-radius $\bar{r}$ lies a proper distance $\ell \,  \textrm{arcsinh}{ }[\bar{r}/\ell]$ from the center of the bubble; the edge of the bubble lies a proper distance $\ell \, \textrm{arcsinh}  [r_0 /\ell]$ from the center of the bubble. Therefore the sphere lies a fraction $\textrm{arcsinh}[\bar{r}/\ell] /\textrm{arcsinh}{ } [r_0/\ell] $ out of the bubble. Here, we plot this fraction as a function of the area-radius of the sphere, $\bar{r}$, for various values of $\ell$. For flat space, this would be a straight line because the area-radius is equal to the proper radius, but for negatively curved spaces the area-radius grows faster than the proper radius. As the space becomes increasingly negatively curved, a sphere with a given area-radius $\bar r$ lies an ever-increasing fraction of the way out; on the plot, the value of the function at a given value of the $x$-coordinate gets ever larger as $\ell$ gets ever smaller. In the limit $\ell \rightarrow 0$ all the foliating spheres accumulate at the edge of the bubble.}
   \label{fig:smoothlimit}
\end{figure}

\begin{enumerate}
\item[3.] { \bf Spheres of nonzero area get expelled from the true-vacuum region.} Consider a sphere of area-radius $\bar{r}$. The distance of this sphere from the boundary of the true-vacuum region is 
\begin{equation}
\textrm{distance from the edge} = \int_{\bar{r}}^{r_0} \frac{dr}{\sqrt{1 + r^2 /\ell^2}} =  \left[ \ell\,  \textrm{arcsinh}  \frac{r}{\ell}\right]_{\bar{r}}^{r_0}  \sim \ell \log \frac{{r_0}}{\bar{r}} .
\end{equation}
This goes to zero in the limit $\ell \rightarrow 0$. The distance from the center of the bubble is
\begin{equation}
\textrm{distance from the center} = \int_0^{\bar{r}} \frac{dr}{\sqrt{1 + r^2 /\ell^2}} =  \left[ \ell\,  \textrm{arcsinh}  \frac{r}{\ell}\right]_0^{\bar{r}} \sim \ell \log \frac{2 \bar{r}}{\ell} .
\end{equation}
This also goes to zero in the limit $\ell \rightarrow 0$, but crucially, it goes to zero \emph{slower} than the distance to the edge
\begin{equation}
\frac{\textrm{distance from the edge}}{\textrm{distance from the center}} = \frac{-1}{\log \ell} \rightarrow 0  .
\end{equation}
All spheres get pushed to the edge, and expelled from the bubble, as shown in Fig.~\ref{fig:smoothlimit}.
\end{enumerate}

\pagebreak

In the limit that the flux-tunneling instanton approaches the bubble of nothing, the six-dimensional volume of the interior of the bubble smoothly goes to zero, the four-dimensional volume of the interior of the bubble smoothly goes to zero, and all spheres of nonzero three- two- or one-volume get smoothly ejected.\footnote{For the 6D Einstein-Maxwell model we have shown that the limit is smooth because while the area-radius of the bubble of  nothing remains nonzero, all foliating spheres get expelled. Another way that the approach to the bubble of nothing can be smooth is for the area-radius of the nucleated bubble to go to zero in the limit. This possibility is realized for some theories, including the 5D Einstein-Maxwell theory.}

\subsection{Bubbles of nothing in stabilized extra dimensions: \\ 
the general case} 
\label{sec:bongeneral}

In the last subsection, we looked at perhaps the simplest  way to perturbatively stabilize extra dimensions---the six-dimensional Einstein-Maxwell model---and argued that the `nothing' that emerges in the bubble of nothing should be thought of as the limit of anti-de Sitter space as the curvature length goes to zero. In this subsection, we will argue that this is not just a feature of the 6D Einstein-Maxwell model, and is in fact true for general compactifications that admit bubbles of nothing.

The three contributions to the Einstein-frame effective potential, $V_4$, in the Einstein-Maxwell model all diverge as the size of the extra dimensions goes to zero, as can be seen in Eq.~\ref{eq:V4radion}  and Fig.~\ref{stableminimum}. The contributions of the flux and the cosmological constant go to plus infinity, and the contribution of the curvature goes to minus infinity. Any radion potential constructed out of such components will necessarily have the property that it diverges as the radion shrinks to zero size: if it diverges to plus infinity then the extra dimensions cannot vanish; if it diverges to minus infinity then our result is established.  We will now show that this remains true even if we take more exotic ingredients than flux, curvature and cosmological constants,  that every compactification that admits a bubble of nothing has the property that $V_4 \rightarrow - \infty$ as the size of the extra dimensions shrinks to zero. 

At first, consider a general compactification that features a bulk stress-energy component that is a perfect fluid and is isotropic in the $m$ extra dimensions. Then we can write the pressure in the extra dimensions as $p = w \rho$, where $w$ is the equation-of-state parameter. 

Let's ask what happens to the higher-dimensional energy density $\rho$ as we shrink the volume of the extra dimensions down to zero size. The continuity equation implies that 
\begin{equation}
\rho \sim (\textrm{Volume})_m^{\,-1 - w}, 
\end{equation}
which is to say that energy density gets more intense with decreasing volume if and only if
\begin{equation}
w>-1. \label{w>1}
\end{equation}
If $w=-1$ then we have a cosmological constant and the higher-dimensional energy density is independent of volume. If $w<-1$ then the higher-dimensional energy density gets less intense as the extra dimensions shrink to zero size (and, conversely, the energy density diverges as the extra dimensions get very large, leading to a `big rip' \cite{Caldwell:2003vq}).  

However, the higher-dimensional energy density is not the quantity that we're interested in. We're interested in the energy density, $V_4$, in 4D Einstein frame. To get from the higher-dimensional energy density $\rho$ to the Einstein-frame energy density, we must first integrate out the extra dimensions, which introduces a factor of $(\textrm{Volume})_m$, and we must then rescale our coordinates to ensure that $M_4$ is independent of the size of the extra dimensions, which as we discussed in Sec.~\ref{sec:fluxbon} 
introduces a factor of $(\textrm{Volume})_m^{\,-2}$. Altogether, then, the Einstein-frame effective potential scales as 
\begin{equation}
{V_4} \sim (\textrm{Volume})_m^{\,-2 - w} ,
\end{equation}
which is to say that the Einstein-frame energy density gets more intense with decreasing volume if and only if
\begin{equation}
w>-2. \label{eq:hardtofind}
\end{equation}
Figure~\ref{fig:moreenergyconditions} shows the form of the effective radion potential for different values of $w$ and $\rho$. 
For $w>-2$ the radion potential vanishes at large volume and diverges at small volume; for $w<-2$ it's the other way around.
\begin{figure}[htbp] 
   \centering \label{fig:moreenergyconditions}
   \includegraphics[width=.8\textwidth]{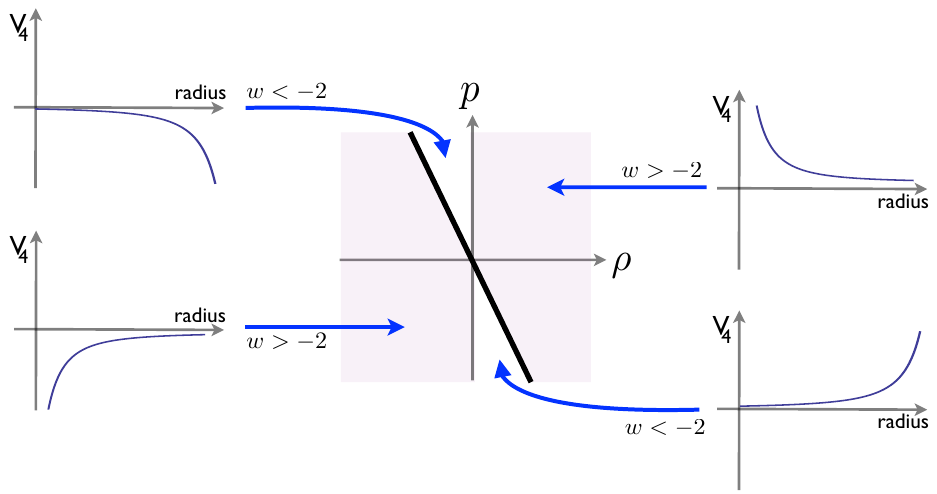}  
   \caption{The radion potential contributions that result from high-dimensional stress-energy sources with various equations of state in the extra dimensions. For $w>-2$ (the shaded regions), the radion potential diverges at small volume and goes to zero at large volume. For $w<-2$ (the unshaded regions), the radion potential vanishes at small volume and diverges at large volume.}
   \label{fig:moreenergyconditions}
\end{figure}

What would the stress-energy ingredients of a perturbative stabilization have to be in order to have bubbles of nothing for which the interior $V_4$ was not going to minus infinity? There are two requirements. First, the stabilization would require at least one $w<-2$ component with positive $\rho$ to force the size of the extra dimensions to zero inside the bubble of nothing. Second, the stabilization would require no immoveable ingredients with $w>-2$; any such ingredient would either positively diverge, forbidding bubbles of nothing, or negatively diverge, sending $V_4 \rightarrow - \infty$. These are two difficult conditions to satisfy, because $w<-2$ components with positive $\rho$ are hard to come by, and $w>-2$ components are hard to exclude. 

That energy components with $w<-2$ are hard to come by is in part a result of the fact that such components must violate the null energy condition, $\rho (1+w) \geq 0$. Moreover, even the standard null-energy-violating ingredients in compactifications\footnote{For example, curvature in the extra dimensions contributes to $G_{\mu \nu}$, but by taking this contribution over to the other side of Einstein's equations we can treat this as an effective contribution to $T_{\mu \nu}$, a contribution which can violate the null energy condition. 
If we have $m$ small dimensions, labeled by $i$, $j$, etc., compactified on a maximally symmetric subspace of curvature $k$ then 
\begin{equation}
R^M_{\, \, N} = \textrm{diag}\{0,0,...,0, k \, g^i_{\, \, j}\} ,
\end{equation}
so 
\begin{equation}
R^M_{\, \, N} - \frac{1}{2} R \, g^{M}_{\, \, N} =  \textrm{diag}\{- \frac{1}{2} m\,  k \, g^{\mu}_{\, \, \nu}, (1-\frac{1}{2} m) \, k \, g^{i}_{\, \, j}\}.
\end{equation}
In other words, curvature makes an effective contribution to the stress-energy with equation of state parameter given by 
\begin{equation}
w = \frac{2-m}{m} > -1.
\end{equation}
Even though positive curvature gives an effective contribution to the stress-energy that violates the null energy condition, extra-dimensional curvature of either sign  gives us an energy component which blows up in the limit of tiny extra dimensions.} do not have \mbox{$w<-2$}. This is related to an argument by Giddings \cite{Giddings:2003zw}.  Giddings showed, following an older argument of Dine and Seiberg \cite{Dine:1985he}, that all of the ingredients in string theory---including the null-energy-violating, the anisotropic, and the imperfectly fluid---all have the property that their contribution to the radion potential goes to zero at large volume.  Correspondingly, their contribution to the radion potential diverges at small volume.

That energy components with $w>-2$ are hard to exclude is exemplified by Casimir energy.\footnote{In addition to Casimir energy, in string theory there will generically be another contribution to the radion potential that diverges at small volume: the closed string tachyon.  Closed strings that wind a small enough cycle on a manifold can develop a tachyonic mass.  In \cite{Adams:2005rb}, it was argued that the endpoint of the condensation of this tachyon is for spacetime to pinch off---`nothing' corresponds to the tachyon rolling all the way down its potential towards $V_4\rightarrow-\infty$  \cite{Horowitz:2005vp,McGreevy:2005ci, Green:2006ku}.}
Casimir energy can make either a positive or a negative contribution to the radion potential depending on the matter content and boundary conditions, but in either case will have $w>-2$. Indeed the presence of Casimir energy even brings the otherwise-exceptional Witten bubble of nothing of Sec.~\ref{sec:wittenbon} into the framework we have been discussing.  If  Casimir energy diverges to plus infinity, as it would for a spinor with periodic boundary conditions, then the bubble of nothing is forbidden. If Casimir energy diverges to negative infinity, as it would for a spinor with anti-periodic boundary conditions, then in this case too we should think of nothing as the limit of anti-de Sitter space as the curvature length goes to zero.

\section{No up-tunneling from nothing}\label{sec:sec3}

In the last section, we discussed how the bubble of nothing instanton mediates the transition, roughly speaking, from something to nothing. In this section, we discuss the reverse process, the transition from nothing to something, the quantum creation of a universe. We show that it is possible to address this issue within the relatively secure foundation of the Coleman-De Luccia formalism \cite{Coleman:1980aw}, which we review in Sec.~\ref{tandu}.  In Sec.~\ref{noquantumcreation}, we argue that it is impossible to up-tunnel using a bubble of nothing instanton. In Sec.~\ref{HTinstanton}, we show that this excludes a number of proposals for the quantum creation of a universe, both the Hawking-Turok instanton \cite{Hawking:1998bn} and the more recent `bubbles from nothing' proposal \cite{BlancoPillado:2011me}.

\subsection{Down-tunneling and up-tunneling}
\label{tandu}

The Coleman-De Luccia (CDL) tunneling prescription is used to calculate the rate of quantum transitions in dynamical spacetimes. This formalism is the most credible part of semiclassical quantum gravity, and the aspiration of this section is to embed questions of the quantum creation of the universe within it. We shall start, in this subsection, by reviewing the CDL formalism, and in particular by highlighting three important elements that will be useful in our upcoming discussion. 
\begin{figure}[h]
  \centering
  \includegraphics[width=5.5in]{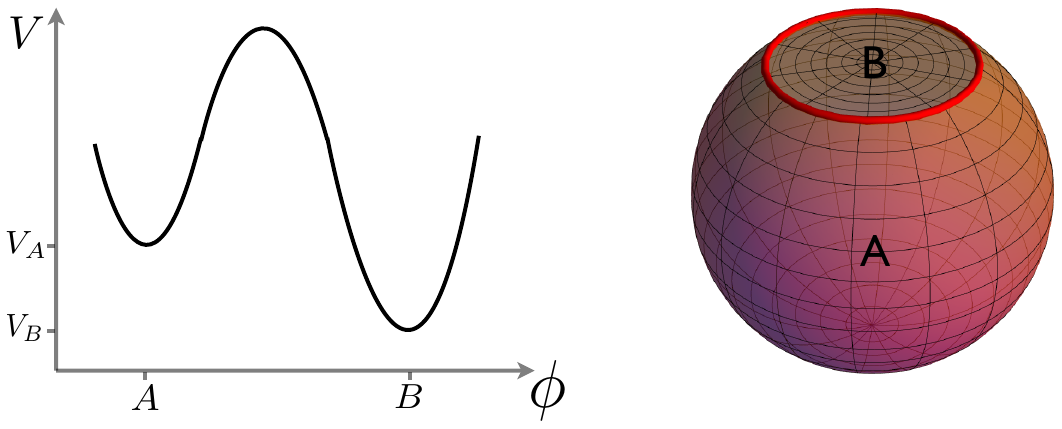} 
  \caption{Because $V_A>0$, the instanton describing down-tunneling from $A$ to $B$ is compact. The instanton has a less-than-horizon-sized bubble of $B$ embedded in $A$. The same instanton also describes the reverse process, up-tunneling from $B$ to $A$.}  \label{fig:uptunnelinstanton}
\end{figure} 

Figure~\ref{fig:uptunnelinstanton}a shows a typical situation for which the CDL formalism is employed. A field uniformly in the $A$ vacuum is classically stable and, through its potential, makes spacetime inflate. Quantum mechanically, however, the field is unstable.  In the semiclassical description, the field and the spacetime make a quantum jump from being uniformly in $A$ to being a less-than-horizon-sized bubble of $B$ embedded in a background of $A$. After the quantum jump, the pressure differential across the wall ensures that the bubble of $B$ grows outwards, so we are left with a region of $B$ that (at least classically) endures. 

The instanton that mediates this decay, shown in Fig.~\ref{fig:uptunnelinstanton}b, is a solution to the Euclidean equations of motion that interpolates between the two vacua.  In the thin-wall limit, it glues together regions of $A$ and $B$ along a thin domain wall; the regions of $A$ and $B$ are sections of Euclidean de Sitter, which is to say they are sections of four-spheres. The CDL prescription gives the tunneling rate from $A$ to $B$,
\begin{equation}
\Gamma_{A \rightarrow B}\sim e^{-\Delta S_\text E/\hbar}, \hspace{.2in} \Delta S_\text E=S_\text{E}(\textrm{instanton}) - S_\text{E}(A).   \label{gammadown}
\end{equation}
The important thing for our purposes is that this rate is nonzero. The instanton is compact, and so has finite action. The Euclidean $B$ vacuum is compact too and so it too has finite action. The exponent in the rate equation, Eq.~\ref{gammadown}, is the difference of two finite quantities and so gives a nonzero tunneling rate. 

But this is not the only transition mediated by the instanton of Fig.~\ref{fig:uptunnelinstanton}b. In fact, the first general principle of CDL tunneling that we wish to highlight is that 
\begin{quote}
\emph{Every instanton describes not one but two transitions.}
\end{quote}
If an instanton calculates the rate for one process, then it necessarily also calculates the rate for the reverse process. In the case at hand, this means that as well as describing `down-tunneling' from $A$ to $B$, the instanton also describes `up-tunneling' from $B$ to $A$ \cite{Lee:1987qc}. A field uniformly in the $B$ vacuum is unstable to the nucleation of a bubble of $A$. After nucleation the bubble wall classically expands into the region of $A$, but the region of $A$ classically endures because there is more than a horizon volume of it, so the de Sitter expansion of the spacetime within the bubble more than compensates for the encroachment of $B$.\footnote{There is a parameter regime when the tension is large for which the instanton features less than a hemisphere of $A$, and hence there is less than a horizon volume of $A$ at nucleation. Nevertheless, the region of $A$ still classically endures, in this case not solely because of de Sitter expansion, but also because of the repulsive gravitational force of the domain wall.}

Both of these processes are described by the same instanton, and the rate to up-tunnel from $B$ to $A$ is given by 
\begin{equation}
\Gamma_{B \rightarrow A}\sim e^{-\Delta S_\text E/\hbar}, \hspace{.2in} \Delta S_\text E=S_E(\textrm{instanton}) - S_E(B) \label{gammaup},
\end{equation}
which differs from Eq.~\ref{gammadown} only in the background subtraction. This background subtraction will turn out to be central to our argument, and arises because $e^{-S_E(B)}$ is the number of microstates in the $B$ macrostate. (The entropy is minus the Euclidean action.) As is required by the principle of detailed balance, the ratio of the tunneling rates, the ratio of Eqs.~\ref{gammadown} and~\ref{gammaup}, is given by $e^{S_E(A) - S_E(B)}$, the ratio of the exponentials of the entropies. 

The essential feature of de Sitter space that allows up-tunneling is that it is a finite system in the sense that it has a finite horizon volume and a finite entropy, and therefore $S_E(B)$ is finite.\footnote{It is the finite volume that is essential for up-tunneling, not the nonzero temperature (though the temperature does speed transitions): finite periodic flat space at zero temperature can up-tunnel, while hot Minkowski cannot \cite{Brown:2011ry}.}  By contrast, when system sizes are infinite, two states can have infinite entropy differences, and then the principle of detailed balance requires that up-tunneling is impossible.  Minkowski space and anti-de Sitter space have infinite horizon volumes and infinite entropies,  which leads to the second general principle that we wish to highlight, that
\begin{quote}
\emph{Up-tunneling is impossible from Minkowski or anti-de Sitter space.}
\end{quote}
Let's see how this fact emerges from the CDL formalism. 

Consider what happens to the up-tunneling transition rate as $V_B$  is lowered towards zero from above. The instanton is only marginally affected: the region of $B$ becomes flatter and smaller, but the instanton stays compact and the Euclidean action stays finite. On the other hand, the Euclidean de Sitter $B$ vacuum is greatly affected: the four-sphere grows without bound, and its action diverges as 
\begin{equation}
S_\text E (V\ge0)=-\frac{24 \pi^2 M_{\textrm{4}}^{\, 4}}{V}.
\end{equation}
As $V_B$ goes to zero, the background subtraction goes to minus infinity so the up-tunneling rate, Eq.~\ref{gammaup}, smoothly vanishes. 
(It's not just that the rate per unit volume, $\Gamma_{B \rightarrow A}$, is going to zero; the expected number of bubbles per Hubble volume per Hubble time, $H_B^{\,-4} \,\Gamma_{B \rightarrow A} $, is also going to zero.  Even though $H_B^{\,-1}$ is diverging in the limit that $V_B$ is lowered to zero, $\Gamma_{B\rightarrow A}$ is going to zero faster---the exponential beats the polynomial.) That this rate is zero is the CDL formalism's way of telling us that up-tunneling from Minkowski space is impossible. 

As $V_B$ is lowered past zero and into negative values, up-tunneling remains impossible. The instanton continues to be compact and continues to have finite Euclidean action, while the Euclidean $B$ vacuum continues to be non-compact and its Euclidean action continues\footnote{Often in AdS/CFT calculations, it is convenient to regulate the Euclidean action of AdS space by adding a canceling divergent boundary term. This convenience does not, however, mean that up-tunneling from AdS is possible! As we are interested in Euclidean action differences between spaces with different asymptotics,  this procedure does not regulate the divergence.} to be infinite
\begin{equation}
S_\text E (V \le0)=-\infty.
\end{equation}
Up-tunneling from AdS to dS is impossible. (Furthermore, up-tunneling from AdS to any higher vacuum is impossible, not just up-tunneling from AdS to dS.)

Notice that up-tunneling did not become impossible because the instanton disappeared or misbehaved. The instanton scarcely changed! Instead, all the action was in the false vacuum subtraction, which blew up as the transition became impossible. There's still a perfectly well-defined instanton describing up-tunneling from AdS to dS, it's just that the rate the CDL prescription assigns to it is zero. This illustrates our third general principle, which is that
\begin{quote}
\emph{Just because a transition is described by an instanton does not mean that the transition is possible.}
\end{quote}
Some transitions described by instantons are assigned zero rate.
(Indeed, when there are two non-positive minima, instantons can exist for which transitions are impossible in \emph{both} directions. This arises when the tension of the corresponding domain wall exceeds the BPS bound---we will discuss this in more detail at the end of the next subsection.)

\subsection{No up-tunneling from nothing}
\label{noquantumcreation}

In Sec.~2 we saw that a bubble of nothing instanton mediates the formation of a hole in spacetime---the decay of something to nothing. In the last subsection, we saw that an instanton that describes one transition necessarily also describes the opposite transition. The question therefore arises: can the bubble of nothing instanton be used to up-tunnel from nothing to something? Does the bubble of nothing instanton mediate the quantum creation of a universe? 

\begin{figure}[h] 
 Ê \centering
   \includegraphics[width=6.2in]{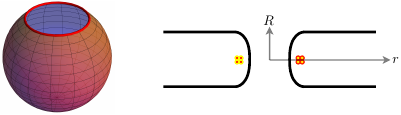}
  Ê Ê \caption{On the left: the instanton describing the formation of a bubble of nothing in de Sitter space.  The region of $B$ from Fig.~10 is replaced by a region of nothing. Recall that, at the lip of the bubble, the extra dimensions pinch off smoothly, as is shown on the right.}
 Ê \label{fig:BubOfNoth}
\end{figure}

In the CDL formalism, the rate to up-tunnel from nothing is given by
\begin{equation}
\Gamma_{\textrm{nothing $\rightarrow$ dS}} \sim e^{-\Delta S_\text E/\hbar}, \hspace{.2in} \Delta S_\text E=S_E(\textrm{instanton}) - S_{\textrm{E}}(\textrm{nothing}),
\end{equation}
and the instanton is shown in Fig.~\ref{fig:BubOfNoth}.

What is the action of nothing? 
Previous authors have implicitly assumed that the action of nothing is zero. ÊThen, when $S_\text{E} (\text{instanton})$ is finite, this leads to the conclusion that up-tunneling from nothing is allowed.  However, there is no need to guess at $S_\text{E} (\text{nothing})$, instead we can appeal to our newfound understanding.

\begin{figure}[htbp] 
 Ê \centering
 Ê \includegraphics[width=4in]{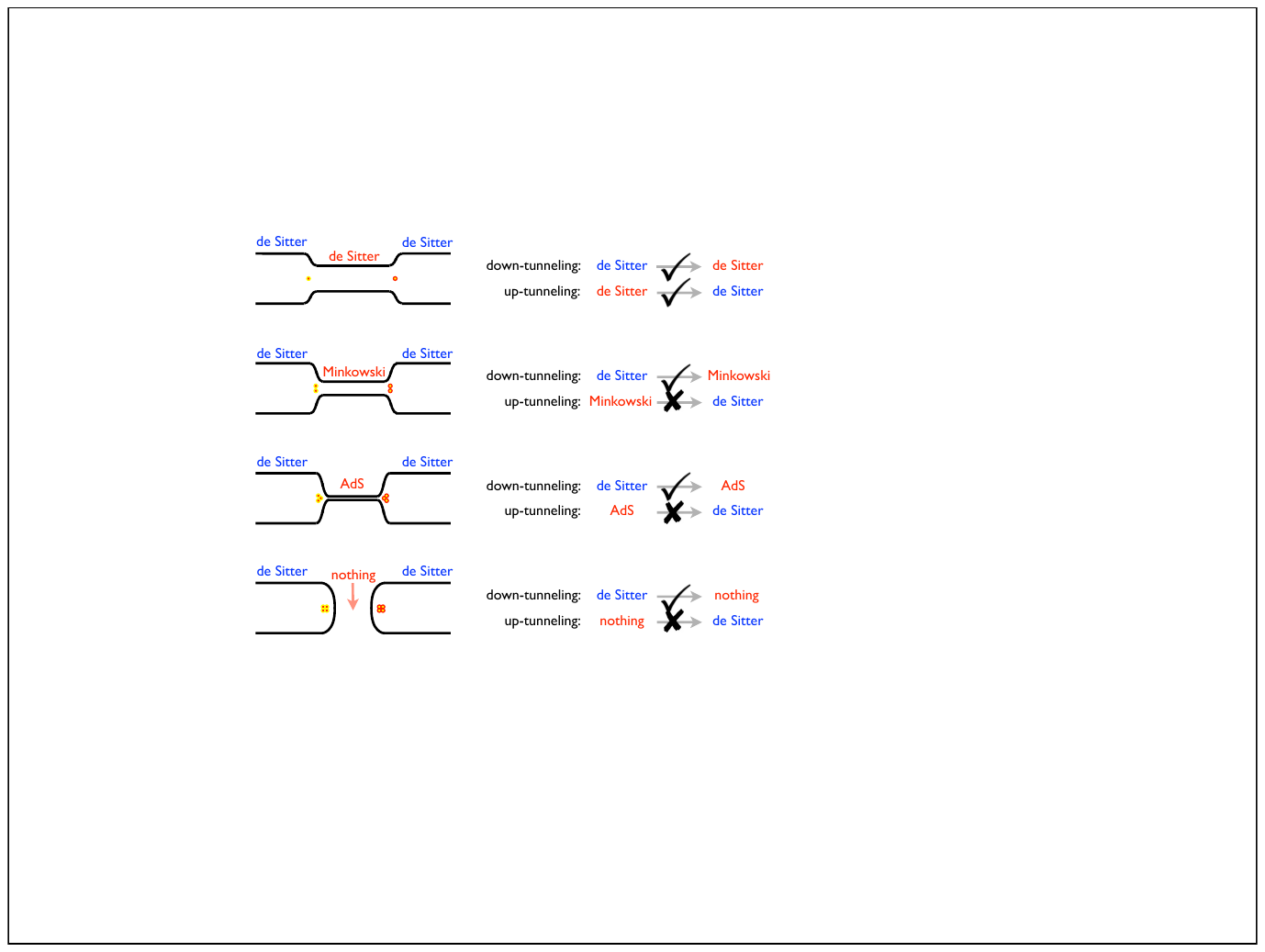} 
 Ê \caption{The four flux-tunneling instantons from Fig.~\ref{fig:fluxsequence}. Each instanton describes two transtions: a down-tunneling transition and an up-tunneling transition. However, as we have seen not every transition described by an instanton is possible; some have zero rate. Beside each instanton we note which of these transitions are possible and which are impossible. Down-tunneling is always possible. Up-tunneling is possible when the interior metric is de Sitter. However, as we move down the family of instantons, up-tunneling is no longer possible once the interior is Minkowski, and remains impossible when the interior is lowered through AdS and on to nothing.}
 Ê \label{fig:possandimposs}
\end{figure}

Let's go back to the simplest case, the 6D Einstein-Maxwell theory. There, in Sec.~\ref{sec:fluxbon}, we saw that  there is a whole family of flux-tunneling instantons. Each of these instantons describes both down-tunneling and up-tunneling transitions but, as we have seen, this does not necessarily mean that these transitions are possible---some of these transitions have zero rate. In Fig.~\ref{fig:possandimposs} we show which transitions are possible, and which transitions are impossible. 

Figure~\ref{fig:possandimposs} shows instantons describing four  decays out of a de Sitter vacuum. The top line shows a decay by just a few units of flux, so that the interior is still a de Sitter vacuum. As a consequence, both down-tunneling and up-tunneling are possible using this instanton. The second line shows a decay by more units of flux, enough that the interior is Minkowski. As a consequence, while down-tunneling is possible using this instanton, up-tunneling is impossible. The third line shows a decay by even more units of flux, enough that the interior is AdS. Again, while down-tunneling is possible using this instanton, up-tunneling is impossible. Finally, the fourth line shows a decay by all the flux, it is a bubble of nothing. Down-tunneling from something to nothing using this instanton is possible, but up-tunneling from nothing to something is impossible. In short, \emph{not only is up-tunneling from nothing a forbidden transition, up-tunneling from nothing is the endpoint of a long sequence of forbidden transitions.}

In the language of Euclidean actions, since nothing should be thought of as a limit of AdS space, and the action of AdS space is minus infinity,
\begin{equation}
S_E(\textrm{nothing}) = - \infty, \textrm{ and so } \Gamma_{\textrm{nothing $\rightarrow$ dS}} = 0.
\end{equation}
Quantum creation of a de Sitter universe is impossible. \\

Quantum creation of a Minkowski or AdS universe is equally impossible. In Sec.~\ref{tandu} we saw that not only is up-tunneling from AdS to dS impossible, up-tunneling from AdS to any higher vacuum is impossible. Since `nothing' is the lowest of all AdS minima, it cannot transition at all. 

A direct consequence is that the `bubbles from nothing' of \cite{BlancoPillado:2011me} (see also \cite{BlancoPillado:2010et,BlancoPillado:2010df}) do not mediate the quantum creation of a universe. `Bubbles from nothing' are Minkowski bubbles of nothing with tension bigger than the BPS-bound, so that the instantons are compact. However, in the CDL formalism  the compactness or non-compactness of an instanton does not affect the question of whether up-tunneling is possible from AdS to Minkowski---either way, it's always impossible. The only thing it does affect is whether the instanton mediates down-tunneling from Minkowski to AdS---for non-compact instantons it does, for compact instantons it does not. That down-tunneling is impossible for compact instantons from Minkowski to AdS is the phenomenon of gravitational blocking described in CDL \cite{Coleman:1980aw}.

\subsection{Bubbles of nothing and the Hawking-Turok instanton}
\label{HTinstanton}

We have argued that the quantum creation of a universe by a bubble of nothing instanton never happens. In this subsection, we show that this means that the quantum creation of a universe via a Hawking-Turok instanton \cite{Hawking:1998bn} also never happens, because the bubble of nothing instanton and the Hawking-Turok instanton are the same thing.

The Hawking-Turok instanton is a four-dimensional instanton with a singularity.  It solves the Euclidean equations of motion for gravity coupled to a scalar field in an inflationary potential; the scalar field is smooth everywhere except at one point---at that point it runs off to infinity in an integrable singularity. 

The bubble of nothing instanton, when described in 4D Einstein frame, is just a Hawking-Turok instanton, as was first realized by Garriga \cite{Garriga:1999xh}.  The radion field in the bubble of nothing plays the role of the scalar field in the Hawking-Turok instanton, and the radion potential becomes the inflationary potential.  The pinch off in the extra dimensions gives rise to a singularity in the scalar field: even though the higher-dimensional metric is smooth, the projection down to four dimensions has a caustic. The fact that the singularity can be smoothly resolved manifests itself as the integrability of the singularity.  In the higher-dimensional theory there is a sphere of minimal area surrounding the origin, but in the coordinates of 4D Einstein frame it appears that this sphere has shrunk to a point. This is related to the conformal rescaling; to preserve the 4D Planck mass, the Einstein-frame rods are infinitely stretched near the pinch-off point, so that zero rods are required to span even the nonzero area. In summary, \emph{the Hawking-Turok instanton is the bubble of nothing instanton written in 4D Einstein frame}.

The Hawking-Turok instanton does not mediate up-tunneling, but not because the singularity means it's not a valid instanton. On the contrary, the singularity is  integrable and the Hawking-Turok instanton is a perfectly legitimate instanton that does mediate a transition: the decay of something to nothing. It doesn't describe the creation of an open universe not because it's not a valid instanton, but because the rate is zero.

\section{Discussion}

We have examined two contexts in which the concept of `nothing'---the absence of spacetime---appears in physics: the bubble of nothing, and the quantum creation of a universe from nothing.  By considering the decays of perturbatively stabilized extra dimensions, we have constructed the bubble of nothing as the limit of a family of otherwise topology-preserving transitions. Every term in the  Einstein-frame radion potential diverges when the extra dimensions get small, so if shrinking to zero size isn't forbidden (which is the case if the potential diverges positively), then shrinking to zero size forces the effective potential to negative infinity.  Therefore the `nothing' inside the bubble of nothing should be understood as the limit of anti-de Sitter space as the curvature length approaches zero.\footnote{The holographic nature of this limit suggests that nothing could be profitably studied in a non-perturbative framework \cite{Balasubramanian:2002am}; since the AdS/CFT dictionary states that $N\sim (\ell_\text{AdS}/\ell_\text{Planck})^{4}$, the dual of nothing appears to be the $N\rightarrow0$ limit of an SU(N) gauge theory.  By giving a limit in which a bubble of nothing is approached by bubbles of AdS, our analysis appears to give a physical realization of \cite{Maldacena:2010un}.}

We then applied this understanding of nothing to the question of the quantum creation of a universe using a Hawking-Turok instanton, which we embedded within the Coleman-De Luccia formalism.  We argued that, since up-tunneling from anti-de Sitter space is forbidden, it is impossible to start with nothing and create a universe.\\ \vspace{.8 cm}

And yet, here we are. How come? Why \emph{is} there something rather than nothing? Let's discuss three possibilities, in ascending order of plausibility.  

First, we showed that up-tunneling from nothing is impossible within the context of the semiclassical low-energy effective theory, but perhaps higher-order terms in the Lagrangian, or quantum corrections to the equations of motion, invalidate our argument. We said that `nothing' is associated with the curvature length going to zero, but when the curvature length gets smaller than the string/Planck scale, there's no reason to expect the corrections to be small.\footnote{That said, corrections are not large for the bubble of nothing itself.  From the higher-dimensional perspective the pinch-off is smooth and the curvature never gets large.  In our way of looking at it, the corrections are small because the volume of the highly curved interior region has gone to zero.}
Maybe these corrections somehow allow up-tunneling from nothing. 


We would find this possibility more plausible if up-tunneling only became impossible as $V_4 \rightarrow-\infty$, if up-tunneling were only forbidden for the final member of the family. ÊHowever, up-tunneling became impossible as soon as $V_4$ reached zero, which is to say as soon as the bubble interior was no longer de Sitter.  There is thus a broad parameter regime for which the solutions are well under control but for which up-tunneling is already forbidden. Ê

Second, in addition to the `nothing' that emerges in the bubble of nothing, from which we have shown tunneling is impossible, perhaps there is a  second, different  type of nothing from which tunneling is possible. This alternative is implicit in the tunneling wavefunction of Vilenkin \cite{Vilenkin:1983xq}, in the Linde tunneling prescription \cite{Linde:1983mx}, and (insofar as it can be thought of as a tunneling from nothing) in the no-boundary proposal of Hartle and Hawking \cite{Hartle:1983ai}. While the existence of the limit-of-AdS type of nothing discussed in this paper is guaranteed by the bubble of nothing, to permit tunneling from nothing these proposals must invoke a second kind of nothing, over and above the `nothing' of the bubble of nothing. 

However, even this is not necessarily enough. ÊThe reason we attach so much significance to tunneling from nothing is that `nothing' seems like a natural place to begin, and so if we can explain how the universe came to tunnel from nothing, we would explain how the universe came to be. But if there are several different kinds of nothing, then we are stuck asking which type is the most likely. For the Hartle-Hawking prescription, the answer seems to be problematic.
The Hartle-Hawking wavefunction assigns amplitudes to three-geometries; if you want to interpret it as tunneling from nothing, then this `nothing' is a three-sphere with zero scale factor. The amplitude the Hartle-Hawking wavefunction assigns to this kind of nothing is small, much less than the amplitude it assigns to a de Sitter universe. ÊThis means that rather than being a natural place for the universe to begin, beginning with this kind of nothing is \emph{less} likely than beginning with a de Sitter universe in the first place---tunneling from nothing has lost any explanatory power. ÊBut even worse than the fact that the Hartle-Hawking wavefunction assigns a very small amplitude to the small-scale-factor nothing, is that it assigns an amplitude to the limit-of-AdS nothing that is \emph{infinite}, so that once the wavefunction is normalized, the probability of being in any other state is zero.
As Hartle and Hawking \cite{Hartle:1983ai} write ``The ground-state wave function obtained by summing over compact four-geometries diverges for large three-geometries in the case $V \leq 0$ and the wave function cannot be normalized.''  This indicates that the ground state of the system is no longer spread out over all the de Sitter vacua, but is instead collapsed into the single lowest vacuum state.  Since the `nothing' of the bubble of nothing is like the most divergent AdS of all, this means that the ground-state wavefunction lies completely in that state, no overlap with any other vacua, and no probability of up-tunneling from nothing. In other words, the existence of any other kind of nothing, besides the `nothing' of the bubble of nothing, seems completely irrelevant: either it has infinitely negative action and you cannot up-tunnel from it, or it has finite action and the Hartle-Hawking wavefunction assigns it an amplitude of zero. 

The third and most likely answer is that we are asking the wrong question. There's so much we don't understand---about the breakdown of the spacetime description at the smallest scales, about quantum gravity, about the ultimate building blocks of existence---that most likely we don't yet even possess the vocabulary to ask a well-posed question. One thing seems clear though: to truly understand everything, we must understand nothing.

\section*{Acknowledgements}
A warm thank you to Jose Juan Blanco-Pillado,  Raphael Bousso, Latham Boyle, Dan Green, Gary Horowitz, Juan Maldacena, Don Page, Ben Shlaer, Eva Silverstein, Paul Steinhardt, Neil Turok, Alex Vilenkin, and Erick Weinberg.

\bibliographystyle{utphys}
\bibliography{mybib.bib}

\providecommand{\href}[2]{#2}\begingroup\raggedright\begin{thebibliography}{10}

\bibitem{Witten:1981gi}
E.~Witten, ``{Instability of the Kaluza-Klein Vacuum},''
  \href{http://dx.doi.org/10.1016/0550-3213(82)90007-4}{{\em Nucl.~Phys.}
  {\bfseries B195} (1982) 481}.

\bibitem{Vilenkin:1983xq}
A.~Vilenkin, ``{The Birth of Inflationary Universes},''
  \href{http://dx.doi.org/10.1103/PhysRevD.27.2848}{{\em Phys.~Rev.} {\bfseries
  D27} (1983) 2848}.

\bibitem{Linde:1983mx}
A.~D. Linde, ``{Quantum Creation of the Inflationary Universe},'' {\em
  Lett.~Nuovo Cim.} {\bfseries 39} (1984) 401--405.

\bibitem{Hartle:1983ai}
J.~Hartle and S.~Hawking, ``{Wave Function of the Universe},''
  \href{http://dx.doi.org/10.1103/PhysRevD.28.2960}{{\em Phys.~Rev.} {\bfseries
  D28} (1983) 2960--2975}.

\bibitem{Hawking:1998bn}
S.~W. Hawking and N.~Turok, ``{Open inflation without false vacua},''
  \href{http://dx.doi.org/10.1016/S0370-2693(98)00234-2}{{\em Phys.~Lett.}
  {\bfseries B425} (1998) 25--32},
\href{http://arxiv.org/abs/hep-th/9802030}{{\ttfamily arXiv:hep-th/9802030}}.

\bibitem{BlancoPillado:2011me}
J.~J. Blanco-Pillado, H.~S. Ramadhan, and B.~Shlaer, ``{Bubbles from
  Nothing},''
\href{http://arxiv.org/abs/1104.5229}{{\ttfamily arXiv:1104.5229 [gr-qc]}}.

\bibitem{Coleman:1980aw}
S.~R. Coleman and F.~De~Luccia, ``{Gravitational Effects on and of Vacuum
  Decay},'' \href{http://dx.doi.org/10.1103/PhysRevD.21.3305}{{\em Phys.~Rev.}
  {\bfseries D21} (1980) 3305}.

\bibitem{Freund:1980xh}
P.~G.~O. Freund and M.~A. Rubin, ``{Dynamics of Dimensional Reduction},''
\href{http://dx.doi.org/10.1016/0370-2693(80)90590-0}{{\em Phys.~Lett.}
  {\bfseries B97} (1980) 233--235}.

\bibitem{BlancoPillado:2009di}
J.~J. Blanco-Pillado, D.~Schwartz-Perlov, and A.~Vilenkin, ``{Quantum Tunneling
  in Flux Compactifications},''
  \href{http://dx.doi.org/10.1088/1475-7516/2009/12/006}{{\em JCAP} {\bfseries
  0912} (2009) 006},
\href{http://arxiv.org/abs/0904.3106}{{\ttfamily arXiv:0904.3106 [hep-th]}}.

\bibitem{Brown:2010mf}
A.~R. Brown and A.~Dahlen, ``{Bubbles of Nothing and the Fastest Decay in the
  Landscape},'' \href{http://dx.doi.org/10.1103/PhysRevD.84.043518}{{\em
  Phys.~Rev.} {\bfseries D84} (2011) 043518},
\href{http://arxiv.org/abs/1010.5240}{{\ttfamily arXiv:1010.5240 [hep-th]}}.

\bibitem{Brown:2010bc}
A.~R. Brown and A.~Dahlen, ``{Small Steps and Giant Leaps in the Landscape},''
  \href{http://dx.doi.org/10.1103/PhysRevD.82.083519}{{\em Phys.~Rev.}
  {\bfseries D82} (2010) 083519},
\href{http://arxiv.org/abs/1004.3994}{{\ttfamily arXiv:1004.3994 [hep-th]}}.

\bibitem{Caldwell:2003vq}
R.~R. Caldwell, M.~Kamionkowski, and N.~N. Weinberg, ``{Phantom Energy and
  Cosmic Doomsday},''
  \href{http://dx.doi.org/10.1103/PhysRevLett.91.071301}{{\em Phys.~Rev.~Lett.}
  {\bfseries 91} (2003) 071301},
\href{http://arxiv.org/abs/astro-ph/0302506}{{\ttfamily
  arXiv:astro-ph/0302506}}.

\bibitem{Giddings:2003zw}
S.~B. Giddings, ``{The Fate of four-dimensions},''
  \href{http://dx.doi.org/10.1103/PhysRevD.68.026006}{{\em Phys.~Rev.}
  {\bfseries D68} (2003) 026006},
  \href{http://arxiv.org/abs/hep-th/0303031}{{\ttfamily arXiv:hep-th/0303031
  [hep-th]}}.

\bibitem{Dine:1985he}
M.~Dine and N.~Seiberg, ``{Is the Superstring Weakly Coupled?},''
  \href{http://dx.doi.org/10.1016/0370-2693(85)90927-X}{{\em Phys.~Lett.}
  {\bfseries B162} (1985) 299}.

\bibitem{Adams:2005rb}
A.~Adams, X.~Liu, J.~McGreevy, A.~Saltman, and E.~Silverstein, ``{Things fall
  apart: Topology change from winding tachyons},''
  \href{http://dx.doi.org/10.1088/1126-6708/2005/10/033}{{\em JHEP} {\bfseries
  0510} (2005) 033}, \href{http://arxiv.org/abs/hep-th/0502021}{{\ttfamily
  arXiv:hep-th/0502021 [hep-th]}}.

\bibitem{Horowitz:2005vp}
G.~T. Horowitz, ``{Tachyon condensation and black strings},''
  \href{http://dx.doi.org/10.1088/1126-6708/2005/08/091}{{\em JHEP} {\bfseries
  08} (2005) 091},
\href{http://arxiv.org/abs/hep-th/0506166}{{\ttfamily arXiv:hep-th/0506166}}.

\bibitem{McGreevy:2005ci}
J.~McGreevy and E.~Silverstein, ``{The tachyon at the end of the universe},''
  \href{http://dx.doi.org/10.1088/1126-6708/2005/08/090}{{\em JHEP} {\bfseries
  08} (2005) 090},
\href{http://arxiv.org/abs/hep-th/0506130}{{\ttfamily arXiv:hep-th/0506130}}.

\bibitem{Green:2006ku}
D.~R. Green, ``{Nothing for Branes},''
  \href{http://dx.doi.org/10.1088/1126-6708/2007/04/025}{{\em JHEP} {\bfseries
  04} (2007) 025},
\href{http://arxiv.org/abs/hep-th/0611003}{{\ttfamily arXiv:hep-th/0611003}}.

\bibitem{Lee:1987qc}
K.-M. Lee and E.~J. Weinberg, ``{Decay of the True Vacuum in Curved
  Space-time},''
\href{http://dx.doi.org/10.1103/PhysRevD.36.1088}{{\em Phys.~Rev.} {\bfseries
  D36} (1987) 1088}.

\bibitem{Brown:2011ry}
A.~R. Brown and A.~Dahlen, ``{Populating the Whole Landscape},'' {\em
  Phys.~Rev.~Lett.} {\bfseries 107} (2011) 171301,
\href{http://arxiv.org/abs/1108.0119}{{\ttfamily arXiv:1108.0119 [hep-th]}}.

\bibitem{BlancoPillado:2010et}
J.~J. Blanco-Pillado, H.~S. Ramadhan, and B.~Shlaer, ``{Decay of flux vacua to
  nothing},'' \href{http://dx.doi.org/10.1088/1475-7516/2010/10/029}{{\em JCAP}
  {\bfseries 1010} (2010) 029},
\href{http://arxiv.org/abs/1009.0753}{{\ttfamily arXiv:1009.0753 [hep-th]}}.

\bibitem{BlancoPillado:2010df}
J.~J. Blanco-Pillado and B.~Shlaer, ``{Bubbles of Nothing in Flux
  Compactifications},''
  \href{http://dx.doi.org/10.1103/PhysRevD.82.086015}{{\em Phys. Rev.}
  {\bfseries D82} (2010) 086015},
\href{http://arxiv.org/abs/1002.4408}{{\ttfamily arXiv:1002.4408 [hep-th]}}.

\bibitem{Garriga:1999xh}
J.~Garriga, ``{Singular instantons and extra dimensions},''
\href{http://dx.doi.org/10.1023/A:1026660116248}{{\em Int.~J.~Theor.~Phys.}
  {\bfseries 38} (1999) 2959--2967}.

\bibitem{Balasubramanian:2002am}
V.~Balasubramanian and S.~F. Ross, ``{The dual of nothing},''
  \href{http://dx.doi.org/10.1103/PhysRevD.66.086002}{{\em Phys. Rev.}
  {\bfseries D66} (2002) 086002},
\href{http://arxiv.org/abs/hep-th/0205290}{{\ttfamily arXiv:hep-th/0205290}}.

\bibitem{Maldacena:2010un}
J.~Maldacena, ``{Vacuum decay into Anti de Sitter space},''
\href{http://arxiv.org/abs/1012.0274}{{\ttfamily arXiv:1012.0274 [hep-th]}}.

\end{thebibliography}\endgroup

\end{document}